\documentclass[iop,revtex4]{emulateapj}
\usepackage{amsmath}

\providecommand{\e}[1]{\ensuremath{\times 10^{#1}}}	
\providecommand{\HII}{H~{\footnotesize II}}			
\providecommand{\HI}{H~{\footnotesize I}}			
\providecommand{\OIII}{[O~{\footnotesize III}]}			
\providecommand{\OII}{[O~{\footnotesize II}]}			
\providecommand{\SII}{[S~{\footnotesize II}]}			
\providecommand{\NII}{[N~{\footnotesize II}]}			
\providecommand{\HA}{H$\alpha$}					
\providecommand{\HB}{H$\beta$}					
\providecommand{\HG}{H$\gamma$}				
\providecommand{\HD}{H$\delta$}					
\providecommand{\HE}{H$\epsilon$} 				
\providecommand{\Te}{T$_e$}						
\providecommand{\cHB}{c$_{\text{H}\beta}$}			
\providecommand{\abun}{12+log(O/H)}				
\providecommand{\LZ}{$L$--$Z$}					
\providecommand{\MZ}{$M$--$Z$}					
\providecommand{\Alecxy}{AGC~198691}			

\slugcomment{11 March 2016 -- Accepted for Publication in ApJ}  

\shorttitle{\Alecxy}
\shortauthors{Hirschauer et al.}

\begin{document}


\title{ALFALFA Discovery of the Most Metal-Poor Gas-Rich Galaxy Known:  AGC 198691}


\author{Alec S. Hirschauer\altaffilmark{1,2}, John J. Salzer\altaffilmark{1,2}, Evan D. Skillman\altaffilmark{3}, Danielle Berg\altaffilmark{4}, \\ Kristen B. W. McQuinn\altaffilmark{5}, John M. Cannon\altaffilmark{6}, Alex J. R. Gordon\altaffilmark{6}, Martha P. Haynes\altaffilmark{7}, Riccardo Giovanelli\altaffilmark{7}, Elizabeth A. K. Adams\altaffilmark{8}, Steven Janowiecki\altaffilmark{9}, Katherine L. Rhode\altaffilmark{1}, Richard W. Pogge\altaffilmark{10,11}, Kevin V. Croxall\altaffilmark{10}, \& Erik Aver\altaffilmark{12}}
\altaffiltext{1}{Department of Astronomy, Indiana University, 727 East Third Street, Bloomington, IN 47405.  {\it e--mail:}  ash@astro.indiana.edu, slaz@astro.indiana.edu, krhode@indiana.edu}
\altaffiltext{2}{Visiting Astronomer, Kitt Peak National Observatory, National Optical Astronomy Observatory, which is operated by the
Association of Universities for Research in Astronomy (AURA) under cooperative agreement with the National Science Foundation.}
\altaffiltext{3}{Minnesota Institute for Astrophysics,School of Physics and Astronomy, 
University of Minnesota, Minneapolis, MN 55455. {\it e--mail:} skillman@astro.umn.edu}
\altaffiltext{4}{Center for Gravitation, Cosmology and Astrophysics, Department of Physics, University of Wisconsin Milwaukee, 1900 East Kentwood Boulevard, Milwaukee, WI 53211. {\it e--mail:} bergda@uwm.edu}
\altaffiltext{5}{University of Texas at Austin, McDonald Observatory, 2515 Speedway, Stop C1402, Austin, TX 78712. {\it e--mail:} kmcquinn@astro.as.utexas.edu}
\altaffiltext{6}{Department of Physics and Astronomy, 
Macalester College, Saint Paul, MN 55105. {\it e--mail:} jcannon@macalester.edu}
\altaffiltext{7}{Center for Astrophysics and Planetary Science, Space Sciences Building, Cornell University, Ithaca, NY 14853. {\it e--mail:} haynes@astro.cornell.edu, riccardo@astro.cornell.edu}
\altaffiltext{8}{ASTRON, the Netherlands Institute for Radio Astronomy, Postbus 2, 7990 AA, Dwingeloo, The Netherlands. {\it e--mail:}  adams@astron.nl}
\altaffiltext{9}{International Centre for Radio Astronomy Research (ICRAR), University of Western Australia, 35 Stirling Highway, Crawley, WA 6009, Australia.  {\it e--mail:} steven.janowiecki@uwa.edu.au}
\altaffiltext{10}{Department of Astronomy, The Ohio State University, 140 West 18th Avenue, Columbus, OH 43210. {\it e--mail:}  pogge@astronomy.ohio-state.edu, croxall.5@osu.edu}
\altaffiltext{11}{Center for Cosmology and AstroParticle Physics, The Ohio State University, 191 West Woodruff Avenue, Columbus, OH 43210.}
\altaffiltext{12}{Department of Physics, Gonzaga University, Spokane, WA 99258. {\it e--mail:}  aver@gonzaga.edu}

\begin{abstract}
We present spectroscopic observations of the nearby dwarf galaxy \Alecxy.
This object is part of the Survey of \HI\ in Extremely Low-Mass Dwarfs (SHIELD) project, which is a multi-wavelength study of galaxies with \HI\ masses in the range of 10$^{6}$-10$^{7.2}$~M$_{\odot}$ discovered by the ALFALFA survey.
We have obtained spectra of the lone \ion{H}{2} region in \Alecxy\ with the new high-throughput KPNO Ohio State Multi-Object Spectrograph (KOSMOS) on the Mayall 4-m as well as with the Blue Channel spectrograph on the MMT 6.5-m telescope.
These observations enable the measurement of the temperature-sensitive \OIII$\lambda$4363 line and hence the determination of a ``direct" oxygen abundance for \Alecxy.
We find this system to be an extremely metal-deficient (XMD) system with an oxygen abundance of \abun\ = 7.02 $\pm$ 0.03, making \Alecxy\ the lowest-abundance star-forming galaxy known in the local universe.
Two of the five lowest-abundance galaxies known have been discovered by the ALFALFA blind \HI\ survey; this high yield of XMD galaxies represents a paradigm shift in the search for extremely metal-poor galaxies.
\end{abstract}



\section{Introduction} 

\indent The gas-phase heavy element abundance of star-forming galaxies is an important physical characteristic that probes enrichment by successive generations of star formation.
Observationally-derived values of metallicities for emission-line galaxies (ELGs) and \HII\ regions provide the anchor points for theoretical models of chemical evolution in galaxies.
The discovery and subsequent study of the lowest metallicity systems, the so-called extremely metal-deficient (XMD) galaxies, has long been recognized as being of special importance.
The XMD galaxies provide some of the best links to the conditions present in the early universe.
They allow us to study star formation and the properties of massive stars in objects that possess inherent similarities to the earliest star-forming systems, as well as providing measurements of primordial abundances as a direct test of Big Bang nucleosynthesis models.
The study of XMDs enables a rare glimpse into key astrophysical processes occurring in the early universe that would otherwise be observationally inaccessible.
\\
\indent An XMD galaxy is classified as one presenting a measured oxygen abundance (used as a proxy for global metallicity) of its interstellar medium (ISM) less than or equal to $\sim$0.1 $Z_{\odot}$, or equivalently \abun\ $\le$ 7.65 (e.g., \citealp{bib:KunthOstlin2000, bib:Kniazev2003, bib:PustilnikMartin2007, bib:Brown2008}).
The most metal-poor galaxies known today have measured nebular oxygen abundances corresponding to roughly 3\% of the Solar value (\abun\ $\sim$ 7.15).
Unfortunately, observations of such extremely low metallicity systems are exceptionally uncommon.
Only four are known in the local universe:
the blue compact dwarf (BCD) galaxy I~Zw~18 \citep{bib:searlesargent1972, bib:Dufour1988, bib:SkillmanKennicutt1993}, the starburst galaxy SBS 0335--052W \citep{bib:izotov1990, bib:Izotov1997, bib:Izotov2005, bib:Izotov2009}, the irregular galaxy  DDO~68 \citep{bib:Pustilnik2005, bib:Berg2012}, and the dwarf galaxy Leo~P \citep{bib:Skillman2013}.
\\
\indent Understanding the astrophysics associated with star formation in extremely low mass and low metallicity systems depends upon the discovery and observational analysis of such systems.
Observationally-determined metallicity relationships such as the luminosity-metallicity (\LZ) and mass-metallicity (\MZ) relations are the consequence of a fundamental correlation between a galaxy's population of stars and the evolution of its metal content:
systems with fewer stars generally exhibit lower heavy-element abundances (e.g., \citealp{bib:Lequeux1979, bib:Skillman1989, bib:Tremonti2004}).
Because robust abundance determinations are observationally challenging to derive for the most intrinsically faint galaxies, and because these systems are harder to find due to their low masses, the current metallicity relations are substantially underpopulated at the extreme low-luminosity end.
In order to provide constraints for theoretical models of chemical evolution that will be representative of {\it both} low-mass and high-mass galaxies, we require additional high-quality metallicity measurements for systems at the lowest luminosities.
\\
\indent There have been ongoing attempts to identify examples of XMDs with measured metallicities lower than  I~Zw~18 (\abun\ = 7.17 $\pm$ 0.04\footnote{\footnotesize  Throughout this paper we list the abundance and emission-line ratios measured for the NW component of I Zw 18.}; \citealp{bib:SkillmanKennicutt1993}) and SBS 0335--052W (\abun\ = 7.13 $\pm$ 0.02; \citealp{bib:Izotov2009}), but these efforts have been largely unsuccessful (e.g., \citealp{bib:KunthSargent1983, bib:KunthOstlin2000, bib:Izotov2012}).
Since these two famous XMDs were originally catalogued in surveys that selected objects that displayed some form of activity, much attention has been focused on searching for galaxies that exhibit the signs of extreme star-formation, such as strong emission lines.
One example of a survey designed to detect large numbers of XMDs is the KPNO International Spectroscopic Survey (KISS; \citealp{bib:Salzer2000, bib:Salzer2001}), a wide-field objective-prism survey that selected objects via H$\alpha$ emission.
Follow-up spectroscopy of over 2400 KISS ELGs identified nearly 1900 star-forming galaxies, with 106 having luminosities fainter than $M_B$ = $-$17.0.
All of these galaxies have metallicity estimates (\citealp{bib:MelbourneSalzer2002, bib:Melbourne2004, bib:Lee2004, bib:Salzer2005b, bib:Hirschauer2015}), yet none have been found to possess  oxygen abundances below \abun\ $\sim$ 7.5.
Surveys like KISS are very effective at finding galaxies with abundances in the vicinity of $\sim$0.1 $Z_\odot$, but have not found any objects in the elusive 0.03 $Z_\odot$ range.   Systematic searches of the Sloan Digital Sky Survey (SDSS; e.g., \citealp{bib:Ahn2012}) spectral database have been somewhat more successful, resulting in a number of identifications of galaxies with \abun\ $<$ 7.5 (e.g., \citealp{bib:Morales-Luis2011, bib:Izotov2012, bib:Guseva2015}) but none with \abun\ $<$ 7.2.
\\
\indent The most recent addition to the XMD galaxies with \abun\ $<$ 7.2 has been Leo~P (\citealp{bib:giovanelli2013, bib:rhode2013, bib:Skillman2013, bib:McQuinnLeoP4, bib:CannonLeoP5, bib:McQuinnLeoP6, bib:McQuinn2015c}).
Leo~P has a measured oxygen abundance of 12+log(O/H) = 7.17 +/- 0.04 (Skillman et al. 2013).  
Despite the similarity of its metallicity to I Zw 18 and SBS 0335-052W, Leo~P is a fundamentally different type of galaxy.  
With a luminosity of M$_B$ = $-$8.8, it is more than 100 times less luminous than these famous BCDs.
Furthermore, Leo~P exhibits a low surface brightness stellar population, in contrast to the compact, high surface brightness nature of I~Zw~18 and SBS 0335--052W. 
Finally, Leo~P was not discovered because of its star-formation activity or strong emission lines.   Instead, it was discovered in a blind \ion{H}{1} survey.
\\
\indent Here we report on spectroscopic observations of the dwarf galaxy \Alecxy\ that have led to the recognition that it is an XMD galaxy with the lowest abundance ever reported for a gas-rich extragalactic source.
Like Leo~P, this object was detected in the Arecibo Legacy Fast ALFA survey (ALFALFA; \citealp{bib:Giovanelli2005, bib:Haynes2011}).  It is included in the extended Survey of \HI\ in Extremely Low-mass Dwarfs (SHIELD; \citealp{bib:Cannon2011}).
ALFALFA has provided the first statistically robust sampling of galaxies that populate the low-mass end of the \HI\ mass function \citep{bib:Martin2010}, including hundreds of objects with M$_{\text{HI}}$ \textless\ 10$^{8}$ M$_{\odot}$.
A subset of these galaxies with 10$^{6}$ M$_{\odot}$ \textless\ M$_{\text{HI}}$ \textless\ 10$^{7.2}$ M$_{\odot}$ and apparent optical counterparts found in the SDSS (\citealp{bib:Abazajian2005}), were selected for the SHIELD project.
The key goal of the latter investigation is to characterize variations in the fundamental galaxy properties as a function of total halo mass.
The SHIELD objects represent some of the lowest \HI-mass galaxies with an obvious stellar population and probable star formation outside the Local Group.
Therefore, based on the standard \MZ\ relationship, they are expected to be among the lowest-metallicity star-forming galaxies in the local universe.
\\
\indent Our observations reveal that \Alecxy\ possesses an oxygen abundance of \abun\ = 7.02 $\pm$ 0.03, thereby earning the distinction of being the lowest metallicity star-forming system known in the local universe.
In \S2 we describe the observational program carried out on this object.
Section 3 details the abundance determination, while \S4 describes the physical properties of the galaxy, compares this galaxy with other known XMDs, and discusses the implications for the discovery of such a low-abundance system.
We summarize our results in \S5.

\section{Observational Program for \Alecxy} 

\begin{figure}
\epsscale{1.1}
\begin{center}
\plotone{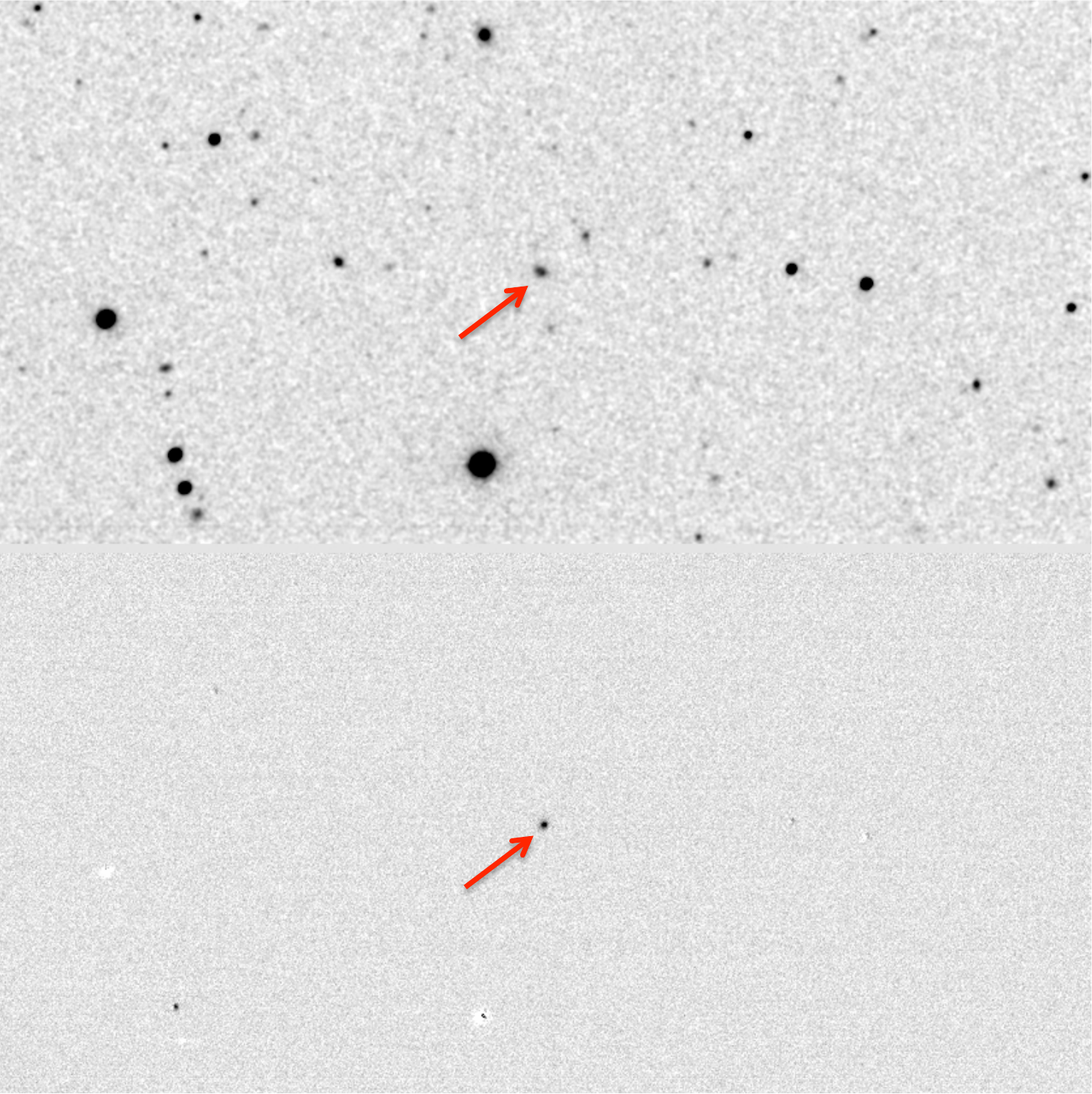}
\end{center}
\caption{\small Broad-band R image (upper) and continuum-subtracted H$\alpha$ image (lower) of the AGC 198691 field obtained with the WIYN 0.9-m telescope.  The field-of-view for both images is 6.0 $\times$ 3.0 arc minutes, with north up and east to the left.  The \ion{H}{2} region detected in the H$\alpha$ image is unresolved in our ground-based data.}
\label{fig:HA_image}
\end{figure}
\indent \Alecxy\ is part of the extended SHIELD sample that was added to the original twelve SHIELD galaxies (e.g., \citealp{bib:Cannon2011, bib:McQuinn2014, bib:McQuinnSHIELD2015}) as additional ALFALFA survey catalogs were completed.
This galaxy has been imaged as part of a Cycle 22 Hubble Space Telescope (\emph{HST})\footnote{\footnotesize Based on observations made with the NASA/ESA \emph{Hubble Space Telescope}, obtained at the Space Telescope Science Institute, which is operated by the Association of Universities for Research in Astronomy, Inc., under NASA contract NAS 5-26555.
These observations are associated with program 13750.} program, and  an \ion{H}{1} map has been obtained with the Westerbork Synthesis Radio Telescope (WSRT)\footnote{\footnotesize The Westerbork Synthesis Radio Telescope is operated by the ASTRON (Netherlands Institute for Radio Astronomy) with support from the Netherlands Foundation for Scientific Research (NWO).}.
Narrow-band H$\alpha$ images were obtained with the WIYN 0.9-m telescope\footnote{\footnotesize The 0.9m telescope is operated by WIYN Inc. on behalf of a Consortium of partner Universities and Organizations (see www.noao.edu/0.9m for a list of the current partners). WIYN is a joint partnership of the University of Wisconsin at Madison, Indiana University, the University of Missouri, and the National Optical Astronomical Observatory.} as part of a larger program to search for \ion{H}{2} regions in the SHIELD sample.
These latter observations revealed a strong, unresolved emission region in \Alecxy\  that provided the target for the spectroscopic observations reported here (Figure \ref{fig:HA_image}).
\\
\begin{figure}
\epsscale{1.3}
\begin{center}
\hspace{-12.0mm}
\plotone{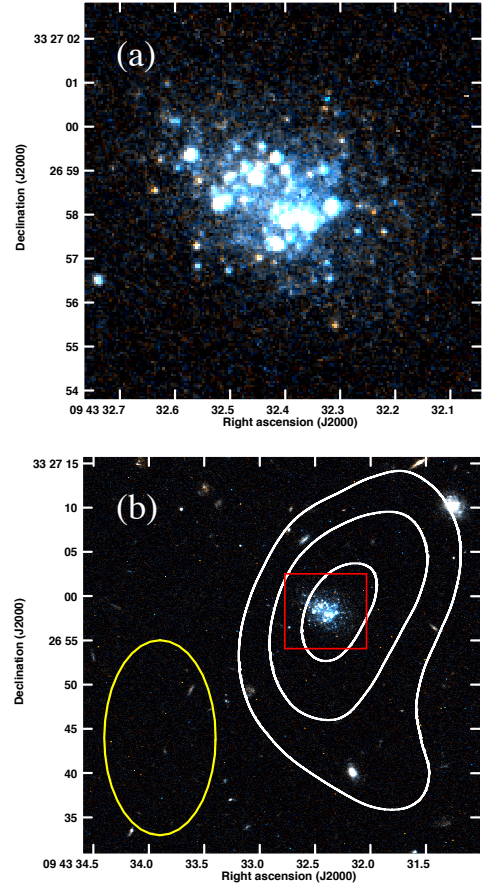}
\end{center}
\caption{\small (a) Hubble Space Telescope (\emph{HST}) image of \Alecxy, created using 20 minute exposures obtained through the \emph{V} (F606W) and \emph{I} (F814W) filters. The galaxy is quite blue, indicating the presence of recently formed massive stars.  The \ion{H}{2} region observed as part of the current program is associated with the clump of bright stars located at the southwestern end of the galaxy.  (b) \ion{H}{1} column density contours from WSRT overlaid on the \emph{HST} image.  The red box denotes the field-of-view of the image shown in (a).  The contours denote column density levels of 2.0, 3.25, and 4.5 $\times$ 10$^{20}$ cm$^{-2}$.
The synthesized beam size is 22 by 12.5 arcsec and is represented by the yellow ellipse.  The \ion{H}{1} distribution in \Alecxy\  is only marginally resolved in our WSRT data.
}
\label{fig:HST_image}
\end{figure}
\indent Figure \ref{fig:HST_image} shows the \emph{HST} image and WSRT \ion{H}{1} map of \Alecxy.
The \emph{HST} observations were taken with the Advanced Camera for Surveys using F606W ($\sim$\emph{V}) and F814W ($\sim$\emph{I}) filters.
The optical appearance of this system indicates a compact structure dominated by luminous blue stars presumably produced in a recent star-formation episode.
The \HA\ emission detected by the WIYN 0.9-m observations is located in the southwest portion of the galaxy.
The WSRT 21-cm radio synthesis map is shown in the lower panel of Figure \ref{fig:HST_image} overlaid on the \emph{HST} image.
The 21-cm emission is only marginally resolved in the WSRT map (beam size of 22 by 12.5 arcsec oriented north-south).
The \ion{H}{1} emission is well-centered on the optical component and exhibits fairly regular column density contours.
There is a hint of an extension to the south, but the reality of this feature is questionable due to the poor resolution and low signal-to-noise of the \ion{H}{1} map.
Forthcoming observations with the VLA should be able to confirm whether this structure is real.
Results from the analysis of both the \emph{HST} and WSRT data will be presented in subsequent papers.   
\\
\indent As part of an ongoing program to obtain spectra of the \ion{H}{2} regions identified in SHIELD galaxies (e.g., \citealp{bib:Haurberg2015}), spectroscopic observations were obtained in April 2015 for \Alecxy\ using the Kitt Peak National Observatory (KPNO) Mayall 4-m telescope utilizing the new KPNO Ohio State Multi-Object Spectrograph (KOSMOS).
Subsequently, we acquired confirming spectra using the MMT 6.5-m telescope\footnote{\footnotesize Observations reported here were obtained at the MMT Observatory, a joint facility of the Smithsonian Institution and the University of Arizona.} and the Blue Channel spectrograph.


\subsection{KPNO 4-m KOSMOS Observations} 

\indent Spectroscopic observations of \Alecxy\ were performed using the KOSMOS spectrograph on the Mayall 4-m telescope at KPNO on 19 April 2015.
Our program utilized both blue- and red-sensitive instrumental setups that provided wavelength coverages of $\sim$3500--6200 \AA\ and $\sim$5000--9000 \AA, respectively.
This enabled the detection of emission lines blueward to \OII$\lambda\lambda$3726,3729 and redward to \SII$\lambda\lambda$6716,6731.
Unfortunately, our red spectral coverage did not quite reach the [\ion{S}{3}] line at $\lambda$9069 \AA; this prevented us from determining a reliable sulfur abundance for \Alecxy.
The instrumental pixel scale is 0.29 arcsec pixel$^{-1}$ in the spatial direction.
The dispersion for the blue spectral setup is 0.66 \AA\ pixel$^{-1}$ while for the red it is 0.99 \AA\ pixel$^{-1}$.
The spectra were taken with a slit width of 1$\farcs$2; the slit extends 10$\arcmin$ in the spatial direction.
The sky conditions were clear at the time of the observations.
An image quality of $\sim$1$\farcs$4 FWHM was derived from the measurement of stellar profiles in the object acquisition images.
\\
\indent \Alecxy\ was observed using the blue setup with three exposures and using the red setup with two exposures, each of which were 900 seconds in length.
The slit was positioned along the parallactic angle as closely as possible in an effort to avoid the effects of differential atmospheric diffraction \citep{bib:Filippenko1982}.
The spectrophometric standard star Feige 34 (\citealp{bib:Massey1988}) was observed to provide calibration data for flux scaling.
Wavelength calibration was accomplished using observations of a HeNeAr comparison lamp.

\subsection{MMT 6.5-m Blue Channel Spectrograph Observations} 

\indent Our initial spectroscopic observations with KOSMOS revealed that \Alecxy\ was exceptionally metal-poor.
However, the large uncertainties in the weak temperature-sensitive \OIII$\lambda$4363 line motivated us to acquire a confirming spectrum.
Additional spectroscopic observations of \Alecxy\ were therefore obtained using the Blue Channel spectrograph on the MMT 6.5-m telescope on 11 and 12 November 2015.
Our program utilized an overall wavelength coverage of 3690--6810 \AA, again enabling detection of emission lines blueward to \OII$\lambda\lambda$3726,3729 and redward to \SII$\lambda\lambda$6716,6731.
The pixel scale along the slit is 0.30 arcsec pixel$^{-1}$ and the 500 lines mm$^{-1}$ grating yielded a dispersion of 1.19 \AA\ pixel$^{-1}$.
Our spectra were taken with a slit width of 1$\farcs$0 on the first night of observations and 1$\farcs$5 on the second night.  The increase in slit width on the second night allowed for the inclusion of more nebular emission.
Seeing values were estimated to be 0$\farcs$8 on both nights.
\\
\indent Five exposures of 1200 seconds were taken of \Alecxy\ on night 1 of the MMT run, with six more obtained on night 2.
The spectrophotometric standard stars BD+28 4211, G191-B2B, and GD 50 \citep{bib:Oke1990} were observed to provide calibration data for flux scaling.
Wavelength calibration was accomplished using observations of a HeNeAr comparison lamp.

\subsection{Data Reduction and Analysis} 

\indent The data reduction for both the KPNO 4-m and MMT 6.5-m observations were carried out with the Image Reduction and Analysis Facility (\texttt{IRAF})\footnote{\footnotesize \texttt{IRAF} is the Image Reduction and Analysis Facility distributed by the National Optical Astronomy Observatory, which is operated by the Association of Universities for Research in Astronomy (AURA) under cooperative agreement with the National Science Foundation (NSF).}.
For both data sets all of the reduction steps mentioned below were carried out on the individual spectral images independently.
The fully reduced 1D spectra were then combined into a single high S/N spectrum for the KPNO run and for each night of the MMT run prior to the measurement stage.

\indent Processing of the two-dimensional spectral images followed standard methods.
The mean bias level was determined and subtracted from each image by using the overscan regions.
A mean bias image was then created by combining 10 zero-second exposures taken on the night of the observations.
This image was subtracted to correct the science images for any possible two-dimensional structure in the bias.
Flat-fielding was achieved using an average-combined quartz lamp image that was corrected for the wavelength-dependent response of the system.
For cosmic ray rejection, we used \texttt{L.A.Cosmic} \citep{bib:vanDokkum2001}, taking special care that no emission lines were ``clipped" by the software.
\\
\indent One-dimensional spectra were extracted using the \texttt{IRAF} \texttt{APALL} routine.
The extraction width (i.e., distance along the slit) was set to 11 pixels or 3.21 arcseconds for the KOSMOS spectra and nine pixels or 2.70 arcseconds for the MMT data.
The extractions were conservatively wide such that modest changes in the extraction width result in negligible changes in the spectra.  
Narrower spectral extractions can achieve higher signal-to-noise in the faintest lines, but at a loss of spectral fidelity.
Sky subtraction was also performed at this stage, with the sky spectrum being measured in regions on either side of the object extraction window.
The HeNeAr lamp spectra were used to assign a wavelength scale, and the spectra of the spectrophotometric standard stars were used to establish the flux scale.
\\
\begin{figure}
\epsscale{1.15}
\plotone{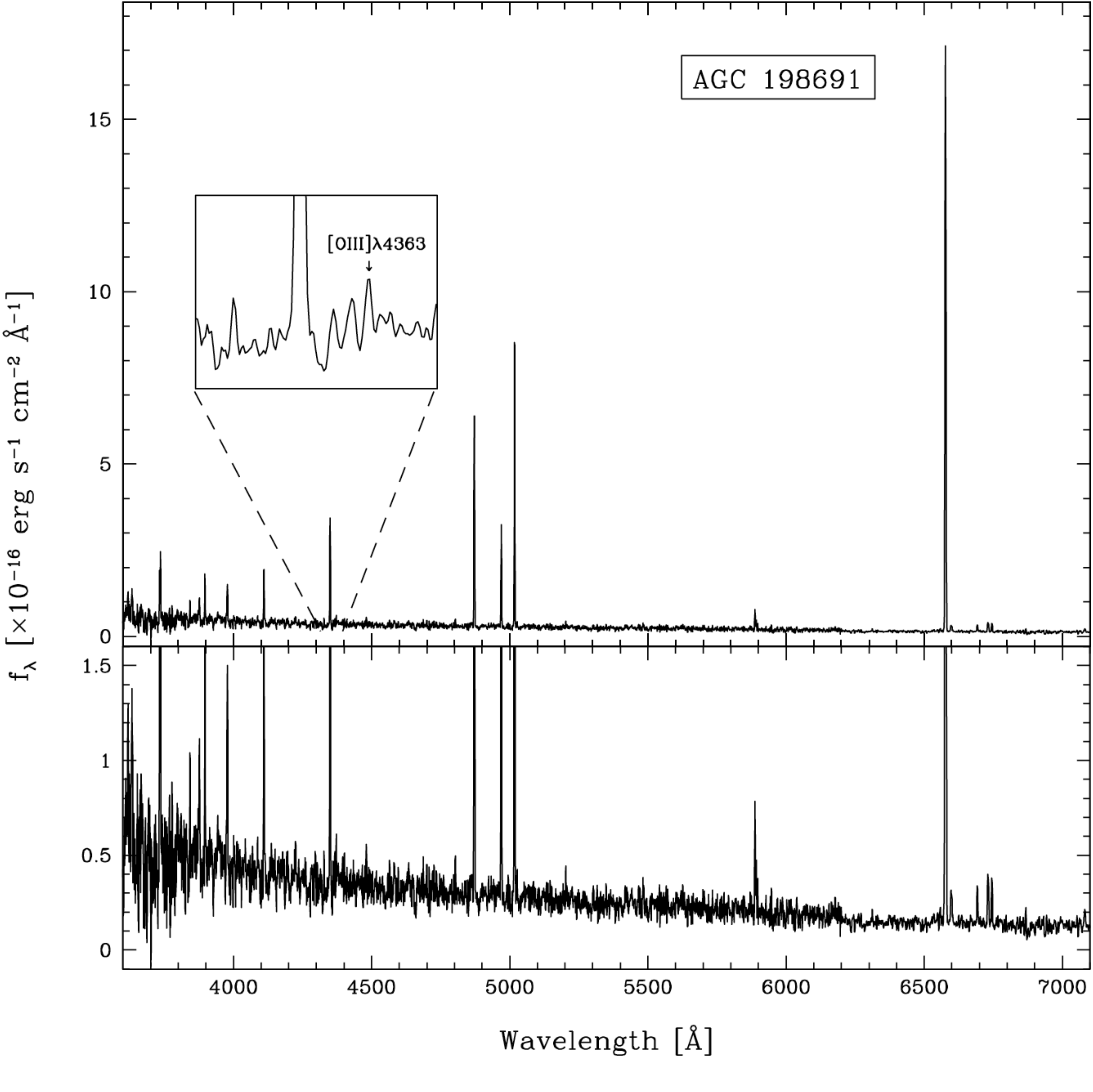}
\caption{\small Optical spectrum of \Alecxy\ (with 3 pixel smoothing) obtained using the KOSMOS spectrograph.
The inset box highlights the \HG\ and \OIII$\lambda$4363 emission lines.  The increased noise between the H$\gamma$ and [\ion{O}{3}]$\lambda$4363 lines is due to the imperfect subtraction of the Hg $\lambda$4358 night sky line.  We stress that the night sky line does not overlap with either of these two nebular lines in our spectra.
Vertical scaling has been reduced on the lower portion of the figure to show increased detail.
}
\label{fig:KOSMOSspectrum}
\end{figure}
\indent The fully reduced KOSMOS spectrum of \Alecxy\ is presented in Figure \ref{fig:KOSMOSspectrum}.
The spectrum represents 45 minutes of integration with the blue setting and 30 minutes with the red setting.
The two independent spectra are plotted together in the figure.
The break point between the red and blue portions is located at $\sim$6200 \AA, and is clearly evident by the change in continuum noise characteristics.
The extraction regions for the red and blue spectra were carefully matched to ensure that the same portion of the galaxy was being extracted.
We checked to ensure that the flux scales in both spectral regions agreed by comparing the fluxes of the [\ion{O}{3}]$\lambda$5007 line that was located in both the red and blue spectra for our instrumental set-up.
The fluxes agreed to within 7\% on average for several objects observed during this run.
We scaled the fluxes of the red spectra to account for this 7\% difference.
The only impact that this flux scaling has on our results for the oxygen abundance in \Alecxy\ is on the determination of the reddening parameter \cHB\ based on the H$\alpha$/H$\beta$ line ratio.
\\
\indent The composite MMT spectrum of \Alecxy\ is presented in Figure \ref{fig:MMTspectrum}.
This final spectrum was created by averaging together the combined spectra taken on the two nights of the MMT observations.
The nightly combined spectra were in turn the averages of the multiple individual spectra taken on the two nights.
This final composite spectrum represents a total of 220 minutes (3.67 hr) of integration time.
\\
\begin{figure}
\epsscale{1.15}
\plotone{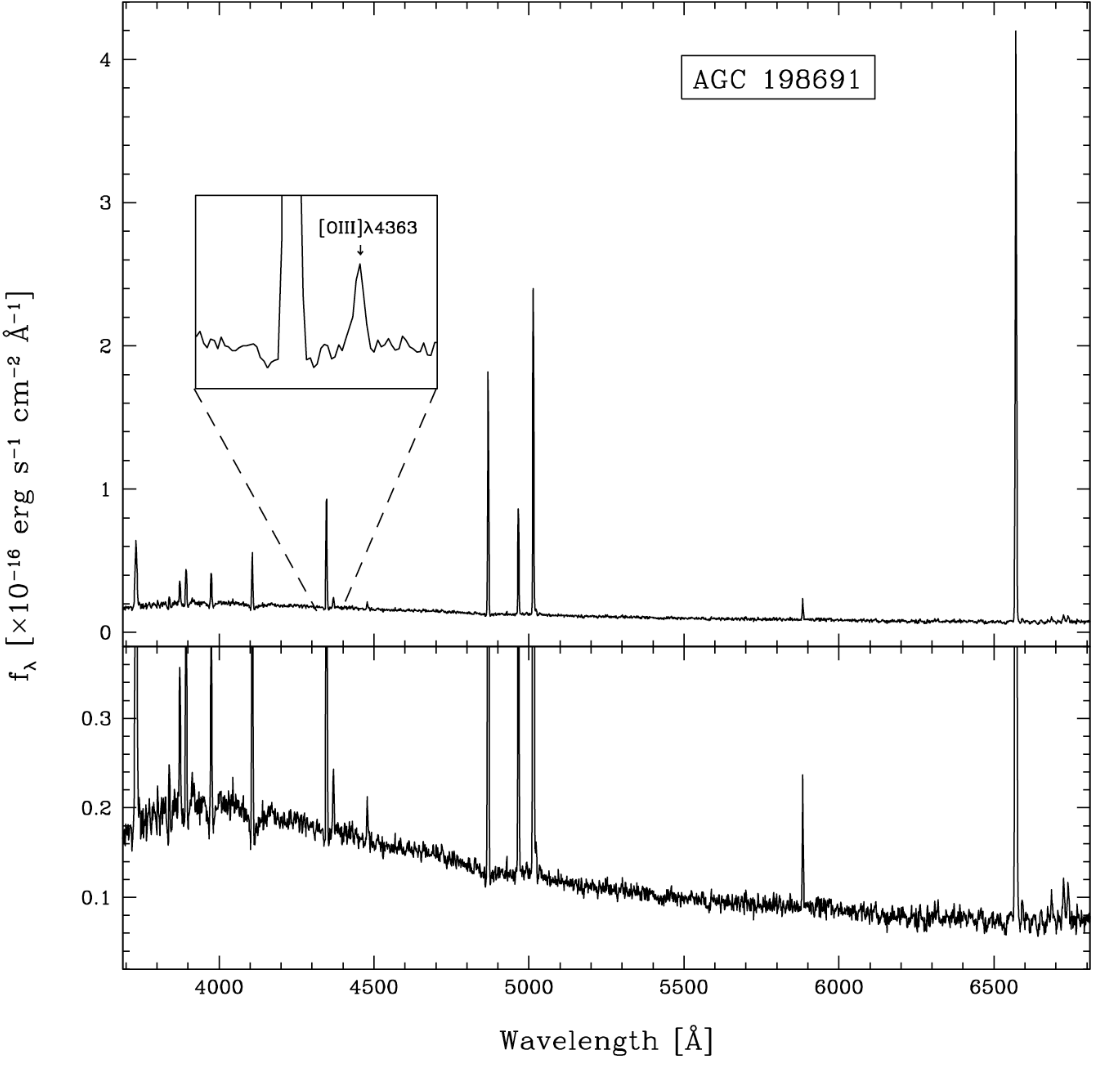}
\caption{\small Composite optical spectrum of \Alecxy\ obtained using the MMT and Blue Channel spectrograph.
The inset box highlights the \HG\ and \OIII$\lambda$4363 emission lines.
Vertical scaling has been reduced on the lower portion of the figure to show increased detail.
}
\label{fig:MMTspectrum}
\end{figure}
\indent For both the KOSMOS and MMT spectra the emission lines were measured using \texttt{SPLOT} by directly summing the flux under each line.  All spectra were measured independently by two of the authors as a consistency check.  In all cases the individual line measurements agreed within the errors. 
\\
\indent For an initial estimate of the internal reddening for \Alecxy\ we calculate \cHB\ from the \HA/\HB\ line ratio.
The value of \cHB\ is then used to correct the measured line ratios for reddening, following the standard procedure (e.g., \citealp{bib:OsterbrockFerland2006}),
\[
\frac{I(\lambda)}{I(\text{H}\beta)} = \frac{F(\lambda)}{F(\text{H}\beta)}\ 10^{-[c_{\text{H}\beta} f(\lambda)]},
\]
\noindent where $f(\lambda$) is derived from studies of absorption in the Milky Way (using values taken from \citealp{bib:Rayo1982}).
The final \cHB\ value used for the reddening correction was determined from a simultaneous fit to the reddening and underlying stellar absorption in the Balmer lines, using all available Balmer line ratios.
Under the simplifying assumption that the same EW of underlying absorption applies to each Balmer line, an absorption line correction was applied to the spectrum that ranged in value from 0 to 5 \AA.
The value for the underlying absorption was varied in 0.5 \AA\  increments until a self-consistent value of \cHB\ for the Balmer line ratios \HA/\HB, \HG/\HB, and \HD/\HB\ was found.
This process led to the determination of a characteristic value for the underlying absorption of 2.0 $\pm$ 0.5 \AA.
\\
\indent Due to the velocity of the source ($V_{\odot}$ = 514 km s$^{-1}$), the Hg $\lambda$4358 night-sky line falls midway between \HG\ and the crucial \OIII$\lambda$4363 temperature-sensitive auroral line (see inset box in Figures \ref{fig:KOSMOSspectrum} and \ref{fig:MMTspectrum}).
The placement of the night-sky line does not interfere with the measurement of either line, but residual flux left over from imperfect  sky subtraction does increase the noise at this location.

\begin{deluxetable*}{cccccc}
\tabletypesize{\scriptsize}
\tablewidth{0pt}
\tablecaption{Emission-Line Intensity Ratios Relative to \HB\ for \Alecxy}
\tablehead{\colhead{Ion}&\colhead{$\lambda$ [\AA]}&\colhead{KPNO 4-m}&\colhead{MMT 6.5-m (night 1)}&\colhead{MMT 6.5-m (night 2)}&\colhead{MMT 6.5-m (composite)}}
\startdata
\noindent	[O II]			&	3727.53		&	0.484	$\pm$	0.027	&	0.4827 $\pm$ 0.0258	&	0.4625 $\pm$ 0.0308	&	0.4680 $\pm$ 0.0246	\\
\noindent	H 9			&	3835.55		&	0.120	$\pm$	0.014	&	0.0627 $\pm$ 0.0054	&	0.0608 $\pm$ 0.0055	&	0.0525 $\pm$ 0.0042	\\
\noindent	[Ne III]		&	3868.74		&	0.141	$\pm$	0.014	&	0.0978 $\pm$ 0.0067	&	0.1074 $\pm$ 0.0079	&	0.0971 $\pm$ 0.0060	\\
\noindent	He I + H 8		&	3888.65		&	0.183	$\pm$	0.015	&	0.1797 $\pm$ 0.0101	&	0.1790 $\pm$ 0.0118	&	0.1683 $\pm$ 0.0090	\\
\noindent	[Ne III] + \HE\	&	3969.56		&	0.162	$\pm$	0.014	&	0.1746 $\pm$ 0.0093	&	0.1839 $\pm$ 0.0114	&	0.1755 $\pm$ 0.0088	\\
\noindent	\HD\			&	4101.74		&	0.250	$\pm$	0.015	&	0.2546 $\pm$ 0.0116	&	0.2618 $\pm$ 0.0146	&	0.2540 $\pm$ 0.0111	\\
\noindent	\HG\			&	4340.47		&	0.468	$\pm$	0.018	&	0.4830 $\pm$ 0.0176	&	0.4873 $\pm$ 0.0233	&	0.4820 $\pm$ 0.0171	\\
\noindent	[O III]			&	4363.21		&	0.044	$\pm$	0.013	&	0.0351 $\pm$ 0.0037	&	0.0428 $\pm$ 0.0037	&	0.0394 $\pm$ 0.0029	\\
\noindent	He I			&	4471.50		&	0.037	$\pm$	0.012	&	0.0288 $\pm$ 0.0035	&	0.0321 $\pm$ 0.0033	&	0.0297 $\pm$ 0.0025	\\
\noindent	\HB\			&	4861.33		&	1.000	$\pm$	0.025	&	1.0000 $\pm$ 0.0296	&	1.0000 $\pm$ 0.0417	&	1.0000 $\pm$ 0.0290	\\
\noindent He I			&	4921.93		&	0.016	$\pm$	0.012	&	0.0100 $\pm$ 0.0031	&	0.0059 $\pm$ 0.0026	&	0.0067 $\pm$ 0.0017	\\
\noindent	[O III]			&	4958.91		&	0.464	$\pm$	0.017	&	0.4172 $\pm$ 0.0135	&	0.4218 $\pm$ 0.0182	&	0.4175 $\pm$ 0.0130	\\
\noindent	[O III]			&	5006.84		&	1.294	$\pm$	0.030	&	1.2687 $\pm$ 0.0373	&	1.2966 $\pm$ 0.0541	&	1.2833 $\pm$ 0.0370	\\
\noindent	He I			&	5875.62		&	0.092	$\pm$	0.013	&	0.0692 $\pm$ 0.0047	&	0.0807 $\pm$ 0.0054	&	0.0750 $\pm$ 0.0041	\\
\noindent	\HA\			&	6562.82		&	2.829	$\pm$	0.058	&	2.7663 $\pm$ 0.1390	&	2.7520 $\pm$ 0.1803	&	2.7583 $\pm$ 0.1374	\\
\noindent	[N II]			&	6583.39		&	0.020	$\pm$	0.016	&	0.0104 $\pm$ 0.0031	&	0.0173 $\pm$ 0.0029	&	0.0161 $\pm$ 0.0021	\\
\noindent	He I			&	6678.15		&	0.021	$\pm$	0.016	&	0.0160 $\pm$ 0.0033	&	0.0298 $\pm$ 0.0035	&	0.0225 $\pm$ 0.0024	\\
\noindent	[S II]			&	6717.00		&	0.038	$\pm$	0.016	&	0.0262 $\pm$ 0.0037	&	0.0391 $\pm$ 0.0041	&	0.0367 $\pm$ 0.0031	\\
\noindent	[S II]			&	6731.30		&	0.025	$\pm$	0.016	&	0.0263 $\pm$ 0.0037	&	0.0342 $\pm$ 0.0038	&	0.0310 $\pm$ 0.0029	\\
\\
\hline
\\
\noindent c(\HB)		& 		& 	0.036 $\pm$ 0.047	&	0.012 $\pm$ 0.054	&	0.073 $\pm$ 0.066	& 	0.040 $\pm$ 0.053	\\
\noindent EW(\HB)		&  		& 	71.7 \AA	&	71.5 \AA &  59.9 \AA  &  64.3 \AA   \\
\noindent F(\HB)*		&  		& 	18.70 $\pm$ 0.30	&  8.31 $\pm$ 0.17  & 10.20 $\pm$ 0.30  &  9.25 $\pm$ 0.19 
\enddata
\tablecomments{* line flux of H$\beta$ in units of 10$^{-16}$ erg s$^{-1}$ cm$^{-2}$}
\label{tab:LineRatios}
\end{deluxetable*}


\section{Abundance Analysis} 

\indent Results of the emission line measurement and analysis, presented as reddening-corrected line ratios relative to \HB, are listed in Table \ref{tab:LineRatios} for the KOSMOS spectrum, the individual night 1 and night 2 spectra from the MMT, and the composite MMT spectrum.
We have opted to measure the lines from the two nights of MMT data separately in addition to measuring the composite spectrum.
This provides a check of our results.
Furthermore, the data obtained on the two nights are not identical, since different slit widths were employed and the telescope pointing may not have been precisely repeated.
The reader is cautioned that the MMT composite spectrum is therefore not necessarily a simple average of two identical spectra.
This is clear in the bottom lines of Table \ref{tab:LineRatios}, where the equivalent width of the H$\beta$ line measured in the night 2 data is reduced relative to the night 1 value because of the broader slit used on that night, while the integrated H$\beta$ line flux is higher.
\\
\begin{figure}
\epsscale{1.15}
\plotone{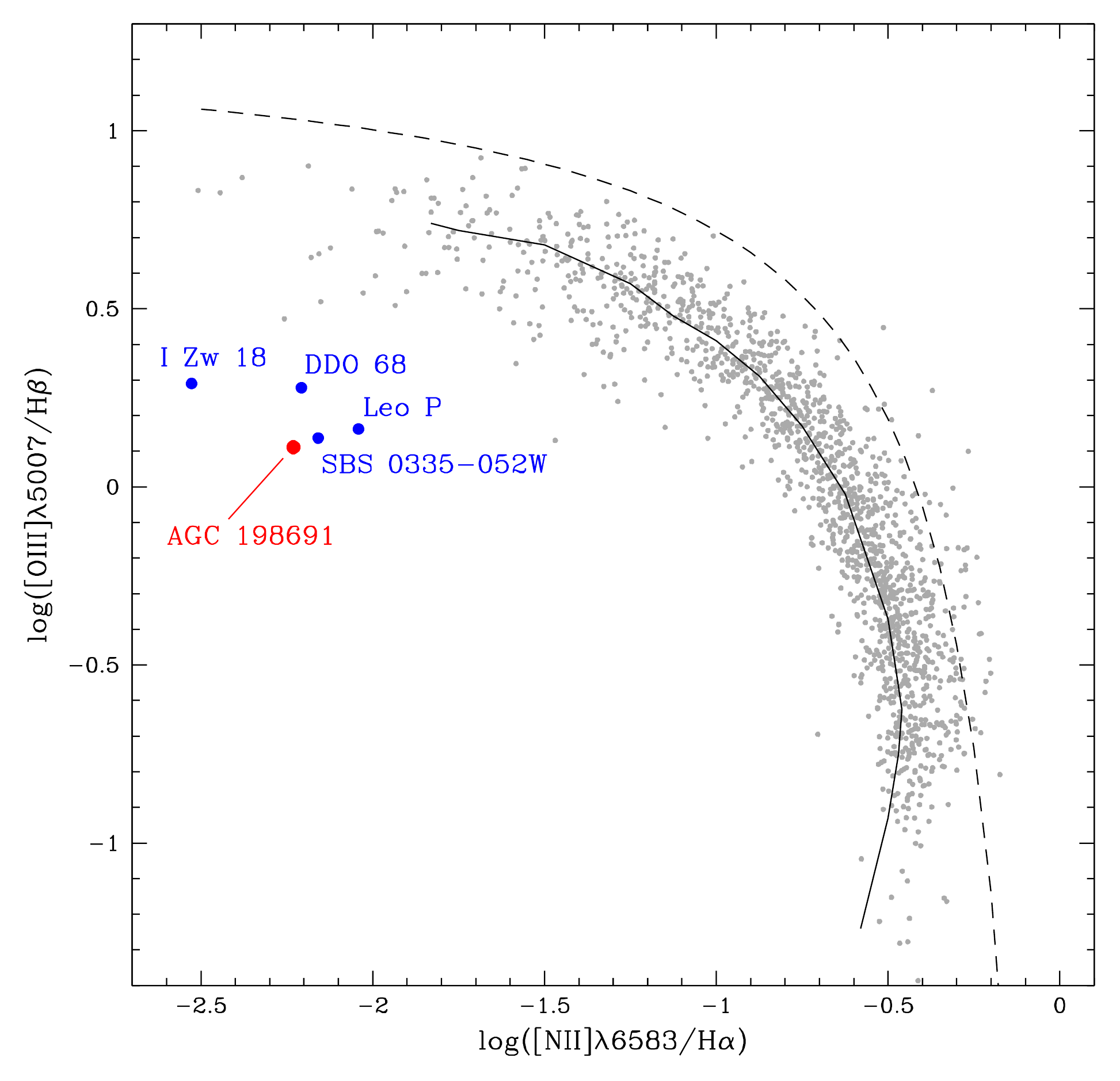}
\caption{\small Spectral excitation diagnostic diagram showing the location of \Alecxy\ (red filled circle).
The four other known low-metallicity star-forming galaxies with \abun\ $\le$ 7.20 are displayed as blue filled circles.
The grey points are star-forming galaxies from the KISS sample, included to provide context (data taken from the references listed in \S 1).
The solid line represents a star formation sequence derived from theoretical models \citep{bib:DopitaEvans1986}, while the dashed line represents an empirically defined demarcation between starburst galaxies and AGN \citep{bib:Kauffmann2003}.
\Alecxy\ is clearly a very low-abundance source.
}
\label{fig:4Dec2015_DD}
\end{figure}
\indent  We show the location of \Alecxy\ in a standard emission-line ratio diagnostic diagram (e.g., \citealp{bib:Baldwin1981}) in Figure \ref{fig:4Dec2015_DD}.
The figure shows \Alecxy\ along with the four previously known XMD galaxies that have \abun\ at or below 7.20 with reliably measured values (uncertainties of 0.05 dex or less).   The line ratios for \Alecxy\ are those from the composite MMT spectrum, while the values for the XMDs come from the references listed in \S 1.
Specifically, the line ratios for I Zw 18 come from \cite{bib:SkillmanKennicutt1993}, for SBS 0335--052W  from \cite{bib:Izotov2009}, for DDO~68  from \cite{bib:Pustilnik2005}, and for Leo~P from \cite{bib:Skillman2013}.
As shown by \cite{bib:Izotov2012} the most metal poor galaxies occupy a restricted location in the traditional diagnostic diagram.
The position of \Alecxy\ indicates that this galaxy has an abundance in the same neighborhood as these other extremely metal-poor systems.
The low value of \NII$\lambda$6584/\HA\ in particular serves as a strong indicator of the exceptionally low metallicity characteristics of \Alecxy.
\\
\indent Calculations of both the temperature and density of the electron gas were carried out using the Emission Line Spectrum Analyzer (\texttt{ELSA}) program \citep{bib:Johnson2006}.  
The electron temperature (\Te) was determined by the oxygen line ratio \OIII$\lambda$4363/\OIII$\lambda\lambda$4959,5007, which exhibits a strong temperature dependence because it arises from two widely separated energy levels in the same ionic species (e.g., \citealp{bib:OsterbrockFerland2006}).
Metallicities derived based on the measurement of the electron temperature are commonly referred to as direct abundances. 
As can be seen by inspection of the inset of Figure \ref{fig:KOSMOSspectrum}, the $\lambda$4363 detection for \Alecxy\ from the KOSMOS data is somewhat weak and noisy (S/N ratio of 3.4), and great care was required in its measurement.
The uncertainty in the final $\lambda$4363 line flux is reflected in the uncertainty in the derived value of \Te.
Our measured value of the electron temperature based on the KOSMOS data is \Te\ = 20440 $\pm$ 3130 K.
For the MMT data the $\lambda$4363 line was more robustly detected, with S/N values ranging from 9.5 to 13.6.  This resulted in a more reliable value for \Te.  Our measured value of \Te\ for night 1 of the MMT run is 18040 $\pm$ 1050 K, for night 2 is 19920 $\pm$ 1060 K, and for the composite MMT spectrum is 19130 $\pm$ 800 K.
We used the [\ion{S}{2}]$\lambda\lambda$6716,6731 doublet to derive values of the electron density.   As with the \OIII$\lambda$4363 line measurement, the weakness of the sulfur lines led to fairly large uncertainties in our derived electron density estimates.  
In the case of the KOSMOS spectrum the [\ion{S}{2}] doublet yielded a flux ratio that placed it in the low-density limit; \texttt{ELSA} assumed an electron density of $n_e$ = 100 cm$^{-3}$.   The three MMT spectra each resulted in electron densities that were in the few-to-several hundred cm$^{-3}$ range.   The derived electron densities and temperatures are presented in Table \ref{tab:Abundances}.
\\
\indent We use the program \texttt{ELSA} to calculate ionic abundances relative to hydrogen.
We estimate the temperature for the low-ionization O$^{+}$ zone by using the algorithm presented in \citet{bib:Skillman1994} based on the nebular models of \citet{bib:Stasinska1990},
\[
t_{e}(\text{O}^{+}) = 2[t_{e}(\text{O}^{++})^{-1} + 0.8]^{-1},
\]
\noindent where $t_{e}$ are temperatures measured in units of 10$^{4}$ K.  
The total oxygen abundance is assumed to be given by
\[
\frac{\text{O}}{\text{H}} = \frac{\text{O}^{+}}{\text{H}^{+}} + \frac{\text{O}^{++}}{\text{H}^{+}},
\]
following the standard practice.
Additional ionization states for other elements that are present in the nebula but do not emit in the optical region of the spectrum are accounted for with ionization correction factors (ICFs).
We use the prescriptions given by \citet{bib:Piembert1969} for the ICFs for N and Ne (see also \citealp{bib:Izotov1994}):
\[
\text{ICF(N)} = \frac{\text{N}}{\text{N}^{+}} = \frac{\text{O}}{\text{O}^{+}},
\]
\[
\text{ICF(Ne)} = \frac{\text{Ne}}{\text{Ne}^{++}} = \frac{\text{O}}{\text{O}^{++}}.
\]
Using the ionic abundances and ionization correction factors given above, we calculate the 
O/H, N/H, N/O, Ne/H, and Ne/O ratios for each data set.  
These values are presented in Table \ref{tab:Abundances}.
Note that we have chosen not to report a sulfur abundance because we have only detected lines from the S$^{+}$ ionization state; sulfur abundances resulting from only a single ionization state are usually deemed to be unreliable.
We have also decided not to present a He abundance based on our current spectra.
This decision was made based on the weakness of a number of the He I lines such as $\lambda$4471, $\lambda$4922, and $\lambda$6876, plus the fact that in the lower dispersion MMT data the $\lambda$5876 line may be contaminated by residuals from the adjacent Na D night sky line.   We plan to use a high-quality follow-up spectrum obtained with the Large Binocular Telescope to carry out an analysis of the Helium abundance of AGC 198691 in the near future.

\begin{deluxetable*}{ccccc}
\tabletypesize{\footnotesize}
\tablewidth{0pt}
\tablecaption{Electron Temperatures \& Densities and Ionic \& Elemental Abundances}
\tablehead{\colhead{Property}&\colhead{KPNO 4-m}&\colhead{MMT 6.5-m (night 1)}&\colhead{MMT 6.5-m (night 2)}&\colhead{MMT 6.5-m (composite)}}
\startdata
\noindent $T_{e}$ (O~{\scriptsize III}) [K] & 20440 $\pm$ 3130 & 18040 $\pm$ 1050 & 19920 $\pm$ 1060 & 19130 $\pm$ 800 \\
\noindent $n_{e}$ (S~{\scriptsize II}) [cm$^{-3}$] & 100 (adopted) & 660 $\pm$ 500 & 340 $\pm$ 260 & 270 $\pm$ 200 \\
\\
\noindent O$^{+}$/H$^{+}$ [$\times$10$^{6}$] & 3.05 $\pm$ 0.34 & 3.75 $\pm$ 0.50 & 2.94 $\pm$ 0.27 & 3.11 $\pm$ 0.24 \\
\noindent O$^{++}$/H$^{+}$ [$\times$10$^{6}$] & 6.97 $\pm$ 0.92 & 8.05 $\pm$ 0.96 & 6.98 $\pm$ 0.51 & 7.35 $\pm$ 0.51 \\
\noindent O/H [$\times$10$^{6}$] & 10.02 $\pm$ 1.20 & 11.80 $\pm$ 1.34 & 9.92 $\pm$ 0.67 & 10.46 $\pm$ 0.68 \\
\noindent 12+log(O/H) & 7.00 $\pm$ 0.05 & 7.07 $\pm$ 0.05 & 7.00 $\pm$ 0.03 & 7.02 $\pm$ 0.03 \\
\\
\noindent N$^{+}$/H$^{+}$ [$\times$10$^{7}$] & 1.10 $\pm$ 0.89 & 0.65 $\pm$ 0.20 & 0.99 $\pm$ 0.16 & 0.95 $\pm$ 0.12 \\
\noindent ICF (N) & 3.29 $\pm$ 0.20 & 3.14 $\pm$ 0.22 & 3.37 $\pm$ 0.20 & 3.37 $\pm$ 0.16 \\
\noindent 12+log(N/H) & 5.56 $\pm$ 0.49 & 5.31 $\pm$ 0.14 & 5.52 $\pm$ 0.08 & 5.51 $\pm$ 0.06 \\
\noindent log(N/O) & -1.44 $\pm$ 0.49 & -1.76 $\pm$ 0.14 & -1.47 $\pm$ 0.08 & -1.51 $\pm$ 0.06 \\
\\
\noindent Ne$^{++}$/H$^{+}$ [$\times$10$^{6}$] & 1.64 $\pm$ 0.33 & 1.39 $\pm$ 0.20 & 1.25 $\pm$ 0.12 & 1.22 $\pm$ 0.11 \\
\noindent ICF (Ne) & 1.43 $\pm$ 0.03 & 1.46 $\pm$ 0.05 & 1.42 $\pm$ 0.04 & 1.42 $\pm$ 0.03 \\
\noindent 12+log(Ne/H) & 6.37 $\pm$ 0.08 & 6.31 $\pm$ 0.06 & 6.25 $\pm$ 0.04 & 6.24 $\pm$ 0.04 \\
\noindent log(Ne/O) & -0.63 $\pm$ 0.05 & -0.76 $\pm$ 0.03 & -0.75 $\pm$ 0.03 & -0.78 $\pm$ 0.02
\enddata
\label{tab:Abundances}
\end{deluxetable*}
%

%
%
\indent The primary results from our observations of \Alecxy\ are seen in column 4 of Table \ref{tab:Abundances}.
The oxygen abundance derived from our initial KOSMOS data is \abun\ = 7.00 $\pm$ 0.05.
The relatively large uncertainty in the electron temperature based on the KOSMOS spectrum caused us to delay reporting our abundance determination for this galaxy until we could obtain a confirming spectrum.
As can be seen in the lower three rows of Table \ref{tab:Abundances}, the MMT spectra do, in fact, reproduce the low abundance implied by the KOSMOS data.
The spectra from the individual nights, as well as the composite spectrum, are all consistent with \Alecxy\ being the lowest abundance galaxy discovered to date.
We adopt the abundance derived from the composite spectrum of 7.02 $\pm$ 0.03 as our final estimate of the oxygen abundance for this system, but assume that subsequent spectroscopic studies of this galaxy will refine this value further.
\\
\indent The nitrogen to oxygen ratio of \Alecxy\ is consistent with other low-abundance sources (e.g., \citealp{bib:vanZeeHaynes2006, bib:Izotov2012, bib:Berg2012}).
The derived value of log(N/O) = $-$1.51 $\pm$ 0.06 from the composite MMT spectrum is slightly higher than the plateau at $-$1.60 defined for XMDs by \citet{bib:IzotovThuan1999}, but still lower than that of Leo~P (= $-$1.36 $\pm$ 0.04; \citealp{bib:Skillman2013}).
\cite{bib:Skillman2013} hypothesized that in the case of Leo~P there has been a period of relative quiescence that has allowed for secondary nucleosynthetic production of nitrogen to build up before additional primary nucleosynthetic production of oxygen took place, thus raising the N/O ratio.
If this interpretation is correct, then the implication for \Alecxy\ is that any similar period of lower star-formation activity would have to be shorter in duration.
This is broadly consistent with its compact, BCD-like appearance (see below).
\\
\indent The ratio of neon to oxygen is expected to be roughly constant between galaxies regardless of their metallicity.
This is because both Ne and O are $\alpha$ elements, which are produced primarily in core collapse supernovae.
Numerous studies have confirmed this result (e.g., \citealp{bib:IzotovThuan1999, bib:vanZee1998, bib:Izotov2012, bib:Berg2012}; see Figure 5 in \citealp{bib:Skillman2013}).
Our value of log(Ne/O) = $-$0.78 $\pm$ 0.02 from the MMT composite spectrum agrees extremely well with the results from these previous studies.
For example, the measured value of log(Ne/O) for Leo~P (\citealp{bib:Skillman2013}) is $-$0.76 $\pm$ 0.03.

\section{Discussion} 

\indent With a measured oxygen abundance of \abun\ = 7.02 $\pm$ 0.03, \Alecxy\ is the most metal-poor star-forming system known in the local universe.
It possesses a metallicity that is 0.11 dex ($\sim$29\%) below that of SBS 0335-052W, the galaxy that had previously been recognized as having the lowest abundance (\abun\ = 7.13 $\pm$ 0.02; \citealp{bib:Izotov2009})\footnote{\footnotesize In \cite{bib:Izotov2009} the authors present two distinct spectral data sets for  SBS 0335-052W.  The highest quality spectrum, obtained with the UVES instrument, represents a single composite spectrum of the galaxy and yields the oxygen abundance quoted here.  This value is consistent with a previous measurement  of 7.12 $\pm$ 0.03 presented in  \cite{bib:Izotov2005}.  The second spectral data set, obtained with the FORS instrument, was spatially resolved into four separate knots, three of which yielded direct abundances of 7.22 $\pm$ 0.07 (their knot \#1), 7.01 $\pm$ 0.07 (knot \#2), and 6.86 $\pm$ 0.14 (knot \#4).  The FORS spectra are lower S/N than the UVES data, resulting in larger uncertainties.   The abundances from the FORS spectra are all within 2$\sigma$ of the UVES measurement, and are consistent with a single, uniform abundance of \abun\ = 7.13.  Future higher S/N ratio spectra of the individual components within SBS0335-052W are needed to establish the presence of possible abundance variations; such observations would be intrinsically interesting and of high value in understanding the true nature of SBS 0335-052W.  
}.
Adopting a solar oxygen abundance of \abun\ = 8.69 (\citealp{bib:Asplund2009}), the metallicity of \Alecxy\ is 0.021 $\pm$ 0.001 Z$_\odot$\footnote{\footnotesize \cite{bib:Steffen2015} have published a new solar oxygen abundance of 8.76.  Using this value results in a slightly lower value for the abundance of \Alecxy\ on the solar scale of 0.018 $\pm$ 0.001 Z$_\odot$.}.  In the following discussion, we explore the properties of \Alecxy\ based on our current knowledge of the system.   

\subsection{Distance and Environment} 

\indent The distance to AGC 198691 is not well constrained at this time.  In this section we present all of the information currently available that  can be used to pin down the distance to this galaxy.  We also discuss briefly the environment within which it resides.
\\
\indent The distance to \Alecxy\ obtained using the velocity flow model employed by the ALFALFA team (\citealp{bib:Masters2005}) is 7.7 Mpc.
This galaxy is located in the direction of the ``local velocity anomaly" (e.g., \citealp{bib:FaberBurstein1988}), so its flow-model distance is likely to be highly uncertain.
If \Alecxy\ is in fact at this distance, it is located within the so-called Leo Spur (e.g., \citealp{bib:Karachentsev2015}), a sparsely populated structure at a distance of roughly 7--10 Mpc.   Since this is the only significant grouping of galaxies along the line-of-sight to \Alecxy\  within a distance of $\sim$20 Mpc, it is tempting to assume that it is in fact located within this structure.

\begin{deluxetable*}{cccccccc}
\tabletypesize{\footnotesize}
\tablewidth{0pt}
\tablecaption{Galaxies Close to \Alecxy\ on the Sky}
\tablehead{\colhead{Name} & \colhead{Velocity} & \colhead{Angular}& \colhead{Minimum} & \colhead{Distance} & \colhead{Distance} & \colhead{Source}\\
& [km s$^{-1}$] & Separation & Projected & [Mpc] &  Method & \\
& & [arcmin] & Separation [kpc] & & &}
\startdata
\noindent 	AGC 198691 &	 514 &    0.0 & --- & --- & --- & --- & \\
\noindent 	UGC 5272     &	 520 &  146.0 &    300 &      7.1 $\pm$ 1.4 & BS & K13 & \\
\noindent 	AGC 205590 &	 494 &  278.7 &    580 &     7.1 $\pm$ 1.4 & TF & K13 & \\
\noindent 	UGC 5427     &	 494 &  365.1 &    820 &    7.69 $\pm$ 0.77 & TRGB & T13 & \\
\noindent 	UGC 5186     &	 549 &    13.0 &       31 &      8.3 $\pm$ 1.7 & TF & K13 & \\
\noindent 	UGC 5209     &	 535 &    75.2 &    230 &  10.42 $\pm$ 0.35 & TRGB & K15 & \\
\noindent 	UGC 5340     &	 507 &  325.2 &  1210 & 12.74 $\pm$ 0.27 & TRGB & C14 & \\
\noindent 	AGC 194100 &	 503 &  222.0 &  1000 &   15.4 $\pm$ 3.1 & TF & K13 & \\
\noindent 	AGC 194054 &	 537 &  217.6 &  1250 &   19.7 $\pm$ 3.9 & TF & K13 & \\
\noindent 	AGC 194068 &	 507 &  235.3 &  1360 &   19.8 $\pm$ 4.0 & TF & K13 & 
\enddata
\tablecomments{Sources for Distances: K13=\cite{bib:Karachentsev2013}, T13=\cite{bib:Tully2013}, K15=\cite{bib:Karachentsev2015}, C14=\cite{bib:Cannon2014}.  Distance Methods: TF=Tully-Fisher Relation, BS=Bright Stars, TRGB=Tip of the Red Giant Branch.   20\% distance errors are assumed for galaxies with TF or BS distances.}
\label{tab:neighbors}
\end{deluxetable*}

\indent It is common practice to look for possible connections between dwarf galaxies and other nearby galaxies that lie along the same line-of-sight. 
There are a number of galaxies with non-velocity-based distances that are located close to AGC 198691 on the plane of the sky that possess comparable redshifts.
In Table \ref{tab:neighbors} we list nine such galaxies that have angular separations of $\sim$6 degrees or less from \Alecxy.
There are four galaxies that have distance estimates of between 7.1 and 8.3 Mpc, consistent with the flow-model distance to \Alecxy.  Unfortunately, only one of these has a reliable tip of the red-giant branch (TRGB) distance.   The others have less accurate distances based on the Tully-Fisher relation or the brightest stars methods.  The galaxy with the smallest angular separation from \Alecxy\ is included in this group: UGC 5186 is located just 13$\farcm$0 away, which corresponds to a minimum separation of 31 kpc if the two galaxies are both located at the distance indicated for UGC 5186.  In fact, UGC 5186 is detected in our WSRT \ion{H}{1} map of \Alecxy.   The other three galaxies in this apparent grouping have minimum separations of between 300 and 820 kpc from \Alecxy, all consistent with the ensemble representing a galaxy group comparable in size to the Local Group at an average distance of 8 $\pm$ 1 Mpc.
\\
\indent The second closest galaxy to \Alecxy\ on the sky is UGC 5209, located 75$\farcm$2 away at a distance of 10.42 $\pm$ 0.35 Mpc from the Milky Way \citep{bib:Karachentsev2015}.  This object appears to be located on the far side of the Leo Spur.  The remaining galaxies in Table \ref{tab:neighbors} have distances beyond 12 Mpc and all have minimum projected separations of 1.0 Mpc or greater from \Alecxy.   One of these, UGC 5340, is also known as DDO 68 which is one of the XMD galaxies referred to in \S 1 ({\abun\  = 7.20 $\pm$ 0.05; \citealp{bib:Pustilnik2005, bib:Berg2012}).   We note the interesting coincidence that \Alecxy\ and DDO 68 are separated by only 5.4$^\circ$ on the sky.
\\
\indent The \emph{HST} imaging data described in \S 2 were obtained as part of a larger effort to ascertain the distances and constituent stellar populations for 18 SHIELD galaxies.
The color-magnitude diagram (CMD) for AGC 198691 has been analyzed following the methodology of \cite{bib:McQuinn2014} with the goal of deriving a TRGB distance.
The short exposure times obtained for this program (one orbit per object split between two filters) result in a relatively shallow CMD.  Thus, our TRGB analysis is unable to yield a reliable distance estimate.
Either AGC 198691 is nearby (distances less than 10-12 Mpc) and its upper RGB is too sparse to yield an accurate TRGB distance, or it is too far away for the TRGB to be detected in our shallow data (distances greater than 12 Mpc).
As we show in \S 4.2, if \Alecxy\ is located at a distance of $\sim$8 Mpc then it likely has a stellar mass comparable to or less than that of Leo~P (\citealp{bib:rhode2013, bib:McQuinnLeoP4, bib:McQuinnLeoP6}).  Consideration of the deep \emph{HST} CMD shown in \cite{bib:McQuinnLeoP6} suggests that a galaxy with a stellar mass significantly lower than Leo~P might well have a RGB that is so sparsely populated that it may not be detectable in our shallow data even at modest distances.  Whether or not \Alecxy\ falls into this category is currently unclear.   At this time we believe that the existing \emph{HST} data cannot rule out a distance to \Alecxy\  as low as 8 $\pm$ 1 Mpc.
\\
\indent To summarize, we feel that one can make a strong case for \Alecxy\ being located within the Leo Spur at a distance of 8 $\pm$ 1 Mpc, based primarily on a group association with UGC 5186, UGC 5272, UGC 5427, and AGC 205590.  This distance estimate may or may not be supported by our shallow \emph{HST} CMD.  The latter cannot immediately rule out any distance in the range of approximately 7 to as high as 20 Mpc.  We note that there are galaxies over this full distance range listed in Table \ref{tab:neighbors}.
\begin{figure}
\epsscale{1.0}
\plotone{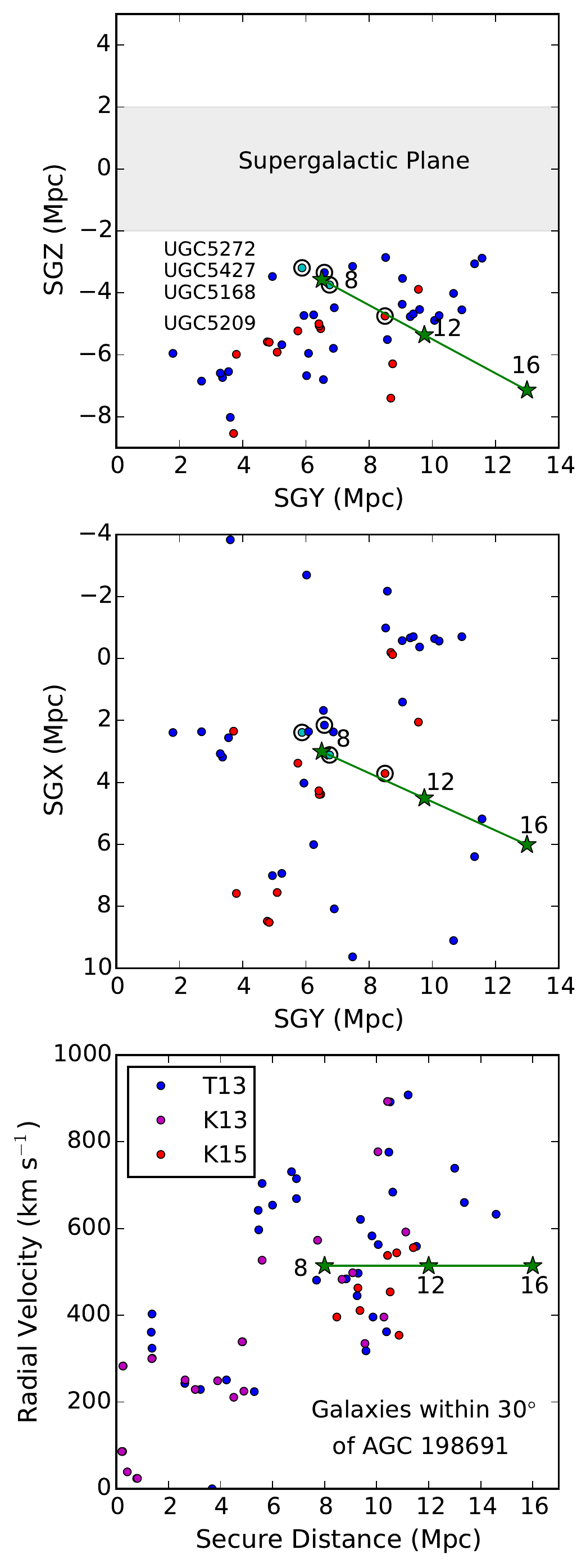}
\caption{\footnotesize The spatial distribution of known galaxies in the direction of AGC 198691.   The top two panels plot the three-dimensional distribution in supergalactic coordinates, while the lower panel shows a Hubble diagram of radial velocity vs. distance.
Red dots are galaxies with distances from \cite{bib:Karachentsev2015}, blue dots represent galaxies from \cite{bib:Tully2013}, and purple dots come from \cite{bib:Karachentsev2013}.
The two cyan points are UGC 5272 and UGC 5186 (labeled in the upper panel) and are discussed in the text.
The green line in all three panels shows the vector along which AGC 198691 lies, with stars indicating distances of 8, 12, and 16 Mpc.
Four galaxies that lie within 1 Mpc of the AGC 198691 position vector are indicated with open circles in the upper two plots.
}
\label{fig:SGplane}
\end{figure}

Figure \ref{fig:SGplane} presents three plots showing the galaxy distribution in this region of space.   The upper two plots illustrate the distribution of galaxies with reliable distances in Supergalactic coordinates, while the lower plot is a simple Hubble diagram and shows how the galaxies within 30$^\circ$ of \Alecxy\  on the sky are distributed in the velocity-distance plane.
For the sake of clarity we only include galaxies in the upper two panels with SGZ less than $-$2 Mpc. 
The green line with stars at 8, 12, and 16 Mpc in all three plots shows the location of AGC 198691 for the expected range of distances discussed above.
Nearly all of the galaxies plotted in Figure \ref{fig:SGplane} have secure distances based on the TRGB method or other primary distance estimates.
The exceptions are UGC 5186 and UGC 5272, two of the galaxies that, along with UGC 5427 and AGC 205590 (not plotted), make up the putative group at 8 $\pm$ 1 Mpc.
The circled points in the upper two panels represent galaxies that have minimum projected separations from \Alecxy\ of less than 1.0 Mpc.
\\
\indent The uncertainty in the distance to AGC 198691 makes it difficult to place this system into the context of its local environment with any certainty.
Nonetheless, it would appear that this galaxy is probably located in a fairly low density region of the nearby universe.
If AGC 198691 is in fact located within the Leo Spur it could still be fairly isolated unless it were in close proximity to one of the galaxies mentioned above.
As seen in Figure \ref{fig:SGplane}, if AGC 198691 is located at a distance of 11 Mpc or beyond then it is extremely isolated (no galaxies within 1 Mpc).
We note that there have been suggestions that low metallicity galaxies are preferentially found in voids (e.g., \citealp{bib:Pustilnik2006, bib:Pustilnik2011}).   If \Alecxy\ followed this trend then distances larger than 11 Mpc would be preferred.
Regardless of its actual distance, it would appear that AGC 198691 ($V_{\odot}$ = 514 km s$^{-1}$) possesses a large negative peculiar velocity and is participating in the same velocity flow as the other galaxies in this region of the local universe \citep{bib:Karachentsev2015}.

\subsection{Physical Properties of AGC 198691} 

\indent In this section we look at the physical characteristics of AGC 198691.
Since the distance is only poorly constrained at this time, we present distance-dependent quantities derived for a range of distances that should bracket the possible values.

\begin{deluxetable*}{cccc}
\tabletypesize{\footnotesize}
\tablewidth{300pt}
\tablecaption{\Alecxy\  -- Observed and Derived Properties}
\tablehead{\colhead{Observed Quantities}&&\colhead{Value}&}
\startdata
\noindent	RA (J2000)				&		&	9:43:32.4			&	\\
\noindent	Dec (J2000)				&		&	+33:26:58			&	\\
\noindent	$m_{V}$					&		&	19.53 $\pm$ 0.03	&	\\
\noindent	$B$--$V$					&		&	0.29	$\pm$ 0.04	&	\\
\noindent	$V$--$R$					&		&	0.14	$\pm$ 0.04	&	\\
\noindent	Angular Diameter [arcsec]		&		&	8.1 $\pm$ 0.2	&	\\
\noindent	F(\HA) [erg s$^{-1}$ cm$^{-2}$]	&		&	(8.49 $\pm$ 0.38)\e{-15}		&	\\
\noindent $V_{\odot}$ [km s$^{-1}$]		&		&	514 $\pm$ 2		&	\\
\noindent $W_{50}$ [km s$^{-1}$]		&		&	33 $\pm$ 2		&	\\
\noindent S$_{\text{HI}}$ [Jy km s$^{-1}$]		&		&	0.53	$\pm$ 0.04		&	\\
\hline
\noindent Derived Quantities			&		&		&	\\
\noindent	Distance (assumed)  [Mpc]	&	8.0			&	12.0	 		&	16.0			\\
\hline
\noindent	M$_{V}$					&	-10.03		&	-10.91		&	-11.53		\\
\noindent	M$_{B}$					&	-9.75			&	-10.63		&	-11.25		\\
\noindent Diameter [pc]				&	320			&	480			&	640			\\
\noindent	L(\HA) [erg s$^{-1}$	]		&	0.64\e{38}		&	1.44\e{38}		&	2.56\e{38}		\\
\noindent	SFR [M$_{\odot}$ yr$^{-1}$]	&	0.00051		&	0.00114		&	0.00202		\\
\noindent \HI\ Mass [M$_{\odot}$]		&	0.80\e{7}		&	1.80\e{7}		&	3.20\e{7}		\\
\noindent	Stellar Mass [M$_{\odot}$]	&	1.6\e{5}		&	3.6\e{5}		&	6.4\e{5}	
\enddata
\label{tab:Parameters}
\end{deluxetable*}

\indent Table  \ref{tab:Parameters} presents relevant observed and derived quantities.
The location of AGC 198691 on the sky places it within the boundaries of the constellation Leo Minor.    For that reason, our group has been referring to it as the {\it Leoncino Dwarf}.
The \emph{BVR} magnitudes and colors are obtained from the SDSS photometry for this galaxy (\citealp{bib:Ahn2012}) using the unpublished conversion relations of R. Lupton.\footnote{\footnotesize see https://www.sdss3.org/dr8/algorithms/sdssUBVRITrans-form.php\#Lupton2005.}
Galactic absorption corrections from \cite{bib:SchlaflyFinkbeiner2011} have been applied for the computation of absolute magnitudes and luminosities.
The uncertainties listed include the formal photometric errors from SDSS as well as the scatter in the regression relations.
The H$\alpha$ flux is measured from our WIYN 0.9-m narrow-band images, which were reduced and calibrated following the procedures described in \cite{bib:VanSistine2016}.
The angular size of the galaxy was measured from our \emph{HST} images and takes into account faint, extended low surface brightness emission that lies beyond the central core containing the bright young stars and the \ion{H}{2} region.
The heliocentric velocity ($V_\odot$), \ion{H}{1} profile velocity width at 50\% of the peak ($W_{50}$), and \ion{H}{1} flux density (S$_{\text{HI}}$) are all from the ALFALFA survey measurements of this source.
\\
\indent We present derived quantities in the lower section of Table \ref{tab:Parameters} for three distances: 8, 12, and 16 Mpc.
We do not quote uncertainties in these parameters since they are dominated by the uncertainty in the distance.
The star-formation rate (SFR) is computed from the H$\alpha$ luminosity using the standard \cite{bib:Kennicutt1998b} conversion factor.
The stellar mass estimates are derived using SED fitting methods as described in \cite{bib:Janowiecki2015b} and Janowiecki et al. 2016 (in preparation).
These values should be taken as being representative only.   Unfortunately, there are currently no NIR flux points for this galaxy which would allow for a more robust stellar mass estimate via the SED fitting method.  AGC 198691 was not detected in either the 2MASS or WISE surveys (e.g., \citealp{bib:2MASS2006, bib:WISE2010}).
Furthermore, the uncertainties due to the presence of strong emission lines in the broad-band filter bandpasses complicates the determination of the stellar mass via this method.
It is worth noting that the value for M$_*$ presented in Table \ref{tab:Parameters} is consistent with estimates obtained using a stellar mass-to-light (M/L) ratio method (e.g., \citealp{bib:BellDeJong2001}).
\\
\indent The color and luminosity of AGC 198691 reveal it to be an actively star-forming dwarf galaxy.
Furthermore, its small size and high surface brightness appearance (Figure \ref{fig:HST_image}) allow us to classify it as a BCD (see \citealp{bib:GildePaz2003}).
The blue absolute magnitude range of $-$9.75 to $-$11.25 makes it one of the lowest luminosity examples of a BCD known.
Most previously identified examples of this class of object possess \emph{B}-band absolute magnitudes in the range $-$13 to $-$17 (e.g., \citealp{bib:JanowieckiSalzer2014}).   
The SFR, while low in an absolute sense, is quite high for a galaxy with this luminosity (e.g., \citealp{bib:Lee2009}).
This is again consistent with our proposed classification as a BCD.
\\
\indent One of the parameters that stands out with AGC 198691 is its very large \ion{H}{1} to stellar mass ratio.
The ratio implied by the values in Table \ref{tab:Parameters} is M$_{\text{HI}}$/M$_*$ = 50, which makes it one of the most gas-rich dwarf galaxies known.
It is worth stressing that this parameter is distance independent; the extreme value we derive will not change as the distance to this source is refined.
For comparison, the M$_{\text{HI}}$/M$_*$ ratio for Leo~P is 2.6 (\citealp{bib:rhode2013}).
Similarly, all of the SHIELD galaxies from the \cite{bib:Cannon2011} sample have M$_{\text{HI}}$/M$_*$ ratios less than two, and most are less than unity \citep{bib:McQuinnSHIELD2015}.  On the other hand, the M$_{\text{HI}}$/M$_*$ value for AGC 198691 is similar to those for the extreme star-forming galaxies I~Zw~18 and SBS0335-052E (M$_{\text{HI}}$/M$_*$ $\approx$ 75 for both; \citealp{bib:Hunt2014}). 
While the M$_{\text{HI}}$/M$_*$ ratio in \Alecxy\ is lower than those of some rare massive systems revealed by ALFALFA (e.g., \citealp{bib:Janowiecki2015}), it is reasonable to hypothesize that the large \ion{H}{1} content of this system may very well play a role in its extreme metallicity.

\subsection{AGC 198691 and the Luminosity-Metallicity Relation} 

\indent The location of \Alecxy\ in a luminosity-metallicity (\LZ) diagram in relation to other XMDs provides a powerful tool for understanding the physics at work with this star-forming system.
Figure \ref{fig:9Dec2015_LZ} shows a set of nearby galaxies with direct abundances taken from the study of \citet{bib:Berg2012} (grey points, her ``Combined Select" sample) plus the five XMD galaxies with \abun\ $<$ 7.20.  
\Alecxy\ is plotted assuming distances of 8, 12 and 16 Mpc. 
The solid line is a linear fit to the data points from the Berg et al. study (her equation (9)). 

\begin{figure}
\epsscale{1.15}
\plotone{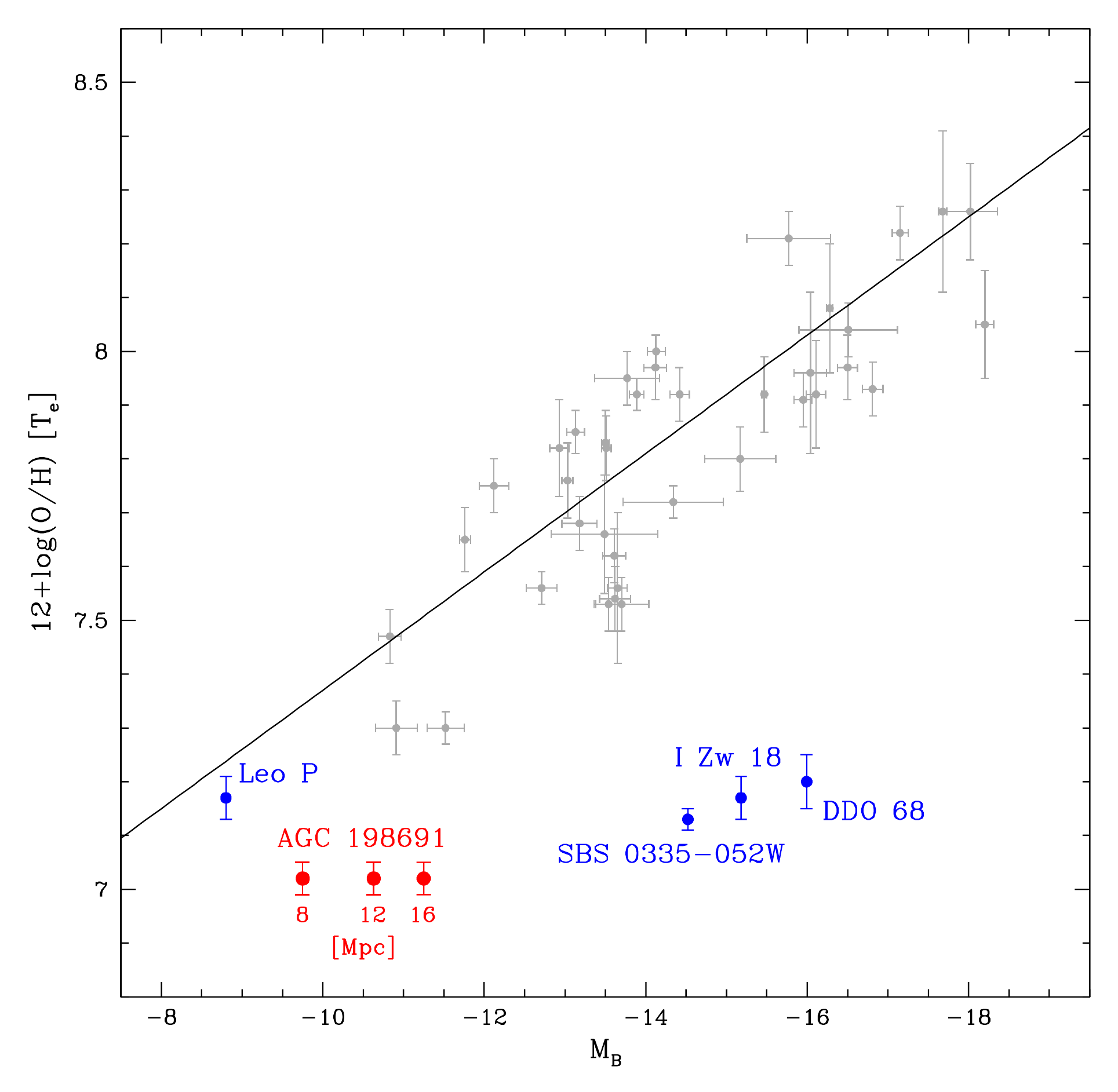}
\caption{\small Luminosity-Metallicity (\LZ) diagram of galaxies with \Te-method abundances taken from \citet{bib:Berg2012} (grey points, her ``Combined Select" sample).
Highlighted is \Alecxy\ (red dots) and the four other known ultra-low-metallicity star-forming galaxies (blue dots).
Because of its uncertain distance, \Alecxy\ is plotted with three different values of M$_B$, corresponding to distances of 8, 12 and 16 Mpc.
The solid line shows the fit to the galaxies from the Berg et al. study.
While the galaxies SBS 0335-052W, I~Zw~18, and DDO~68 fall considerably under the line, Leo~P and \Alecxy\ are more consistent with the extrapolation of the  trend defined by the Berg et al. galaxies, albeit at significantly lower abundance and luminosity.
}
\label{fig:9Dec2015_LZ}
\end{figure}

\indent The exceptionally metal-poor systems SBS 0335-052W, I~Zw~18, and DDO~68 all lie substantially \emph{below} the fit to the rest of the data.
As discussed by \cite{bib:EktaChengalur2010} this may imply that the star-formation mechanism at work in these systems is atypical, being affected by the infall of significant amounts of pristine gas.
The infall could both drive the observed star-formation event as well as dilute the gas resulting in under-enrichment relative to their luminosities.
In contrast, the locations of Leo~P and \Alecxy\ lie far closer to the extrapolation of the fit line defined by the galaxies of the Berg et al. sample.
This might be interpreted as an indication that these two systems are forming stars in a manner more similar to expectation for typical dwarf irregular galaxies.
The fact that \Alecxy\ does exhibit a modest offset from the extrapolated fit line as well as from the relatively quiescent Leo~P could be due to its BCD-like nature (i.e., a modest enhancement in the B-band luminosity moves it to the left in the diagram).
Alternatively, the gas-rich nature of this object might be an indication that it too has experienced recent gas infall that has reduced its observed metallicity.
In either case, the impact of these putative evolutionary mechanisms is clearly far less dramatic than whatever has impacted the three more luminous XMD galaxies shown in Figure \ref{fig:9Dec2015_LZ}.
The determination of an accurate distance to \Alecxy\ would provide valuable insight into the possible mechanisms at play, since it would allow for a much better understanding of the environment within which this system resides.

\subsection{Searching for XMD Galaxies} 

\indent As mentioned in \S1 of this paper, there have been significant efforts made to detect XMD galaxies through the years.
A large fraction of these efforts focused on discovering and observing star-forming dwarf galaxies with the strongest levels of activity (e.g., I~Zw~18 and SBS 0335-052W).
If the premise of \cite{bib:EktaChengalur2010} that these XMDs are strongly affected by infall of pristine gas is correct, we can now appreciate why these objects are so rare.
Existing surveys like KISS could not have failed to discover additional examples of objects like I~Zw~18 and SBS 0335-052W if they were present in the nearby universe with even modest volume densities.
The implication is that the physical processes that shape galaxies like I~Zw~18 and SBS 0335-052W occur only rarely.
It is also likely that the strong starbursts that are generated by the proposed gas infall produce sufficient amounts of new metals that the observed abundances increase back to ``normal" levels fairly quickly.
\\
\indent It is worth noting that \Alecxy\ is located in the area of the sky covered by the 3rd installment of the Case Low-dispersion Northern Sky Survey (\citealp{bib:PeschSanduleak1986}).
The Case survey utilized objective-prism spectra to select galaxies that had detectable emission lines (usually [\ion{O}{3}]), blue continua, or both.
Most of the objects cataloged in the Case survey were brighter than $m_{B}$ = 18, suggesting that \Alecxy\ ($m_{B}$ = 19.8) was simply too faint to be detected by the survey.  
\\
\indent A second method for searching for XMDs has been to obtain spectra of \ion{H}{2} regions in nearby gas-rich dwarf irregular galaxies (e.g., \citealp{bib:Skillman1988, bib:Skillman1989, bib:Skillman1989b, bib:Skillman2003, bib:vanZeeHaynes2006, bib:Berg2012, bib:James2015}).
The existence of the \LZ\ relation implies that the observation of \ion{H}{2} regions in the lowest luminosity galaxies should result in the discovery of numerous XMDs.
However, the shortcomings of this approach are readily evident in  Figure \ref{fig:9Dec2015_LZ}.
The \LZ\ relation shown (solid line) only reaches \abun\ of less than 7.20 for luminosities fainter than $M_B$ $\sim$ $-$9.
Galaxies with this level of luminosity have only been discovered in or very near to the Local Group.
Furthermore, their low masses imply that even if they have retained an appreciable gas reservoir, the probability of catching them with current star formation is probably fairly low.
The obvious example here is Leo~P, which falls very close to the regression line in the \LZ\ diagram.
Leo~P contains a single \ion{H}{2} region that is ionized by a single O star (\citealp{bib:rhode2013, bib:McQuinnLeoP6}).
If Leo~P were observed in 5-10 million years would it possess an \ion{H}{2} region that would allow a derivation of its metal abundance?
\\
\indent This second method of trying to detect XMDs can be effective but is mainly limited by our ability to catalog large numbers of low-luminosity, low-mass galaxies in the nearby universe.
The work to date has been extremely effective at filling in the \LZ\ diagram at luminosities brighter than $M_B$ $\sim$ $-$13, but has resulted in the measurement of abundances that are typically higher than \abun\ = 7.5.
\\
\indent The fact that the two most recently discovered XMDs, Leo~P and \Alecxy, were both discovered in the ALFALFA blind \ion{H}{1} survey immediately suggests a third approach for searching for low-metallicity galaxies.
The extreme sensitivity of the ALFALFA survey has allowed for the detection of hundreds of low \ion{H}{1} mass galaxies beyond the Local Group (e.g., \citealp{bib:Martin2010}).
Naturally, any dwarf galaxy contained within the ALFALFA catalog will have the raw materials necessary for forming stars (i.e., cold gas).
The SHIELD project \citep{bib:Cannon2011} was initiated to study in detail ALFALFA detections with the lowest \ion{H}{1} masses that also possess a clear stellar population.
In the original SHIELD sample, nine of the twelve galaxies are fainter than $M_B$ = $-$13, which puts them into the regime of being interesting targets for possible XMDs.
\\
\indent A first attempt to derive metal abundances for the initial sample of SHIELD galaxies is described in \cite{bib:Haurberg2015}.
Only eight of the twelve galaxies in \cite{bib:Cannon2011} had H$\alpha$ fluxes strong enough to make them viable spectroscopic targets.
Of these, two yielded direct measurement abundances, while for the remaining six strong-line methods were used to estimate O/H.
One SHIELD galaxy was found to have an abundance of \abun\ = 7.41, while the remaining seven had \abun\ between 7.67 and 8.08.
In other words, the abundances measured for the original SHIELD galaxies are similar to those measured for other nearby gas-rich dwarf irregulars.
\Alecxy\ was observed with KOSMOS on the KPNO 4-m as part of the second round of SHIELD galaxy spectroscopy.
The same observing run has yielded several additional metallicity estimates, including two with \abun\ between 7.3 and 7.5 (Hirschauer et al., in preparation).
Many additional low \ion{H}{1} mass targets remain to be observed.
The use of the ALFALFA survey to pre-select low luminosity dwarf galaxies as targets for follow-up spectroscopy is a remarkably fruitful approach to expanding the number of known XMD galaxies.

\section{Summary} 

\indent We have presented KPNO 4-m and MMT 6.5-m spectroscopic observations of the dwarf irregular galaxy \Alecxy\ (a.k.a., the {\it Leoncino Dwarf}), which was discovered in the ALFALFA survey and included in the SHIELD sample of low \ion{H}{1} mass systems.
The analysis of our spectral data has resulted in the derivation of a ``direct-method" oxygen abundance of \abun\ = 7.02 $\pm$ 0.03.
{\bf This metallicity makes \Alecxy\ the most metal-poor galaxy known in the local universe}.
The oxygen abundance of this system is a full 0.11 dex below that of SBS 0335-052W and 0.15 dex below I~Zw~18.
While the spectra in the current study were not suitable for providing an estimate of the helium abundance in this galaxy, we suggest that future observations would very likely be fruitful in yielding an estimate of the primordial helium abundance.
\\
\indent \Alecxy\ is a small, high surface brightness system that we classify as a blue compact dwarf galaxy.
It is one of the lowest luminosity members of the BCD class ever discovered.
Its observed color of $B$--$V$ = 0.29 is indicative of a galaxy dominated by a young stellar population.
A secure distance to this system is not currently available.
Our analysis of the available data appears to constrain the distance to be within the range of 7--16 Mpc.
If the distance is in the nearer portion of that range (7--11 Mpc) it is located in the sparsely populated Leo Spur.
If it has a distance in the farther portion of this range (11--16 Mpc) then it is likely to be quite isolated, with no neighbors within 1 Mpc.
The determination of an accurate distance for this object is a high priority.
Regardless of its distance, it is clear that \Alecxy\ is extremely gas rich.
\\
\indent The recent discoveries by the ALFALFA survey of two new XMD systems with \abun\ less than 7.20 speak to the power of the \ion{H}{1} selection method for searching for metal poor systems.
Both Leo~P and \Alecxy\ are extreme systems within the Local Supercluster that remained unrecognized until their detection in the blind \ion{H}{1} survey carried out at Arecibo.
Hundreds more ALFALFA detections remain to be explored.



\acknowledgements

The work presented in this paper is based in part on observations obtained at Kitt Peak National Observatory, National Optical Astronomy Observatory (NOAO Proposal ID 2015A-0408, PI: J. Salzer), which is operated by the Association of Universities for Research in Astronomy (AURA) under cooperative agreement with the National Science Foundation.  Observations were also obtained at the MMT Observatory, a joint facility of the Smithsonian Institution and the University 
of Arizona. MMT observations were obtained as part of the University of Minnesota's guaranteed time on Steward Observatory facilities through membership in the Research Corporation and its support for the Large Binocular Telescope.    
We gratefully acknowledge the support of the staffs of the KPNO and MMT Observatories for their expert assistance during the course of our observing runs, and would also like to thank the anonymous referee for several useful comments and suggestions.
JJS and ASH received financial support from the College of Arts and Sciences of Indiana University that helped to enable this work.
The ALFALFA team at Cornell is supported by NSF grants AST-0607007 and AST-1107390 to RG and MPH and by a grant from the Brinson Foundation.
JMC is supported by NSF grant AST-1211683.
This work was supported in part by NASA through grant GO-13750 from the Space Telescope Institute, which is operated by AURA, Inc., under NASA contract NAS5-26555.


\bibliographystyle{apj}
\bibliography{Bibliography}

\end{document}